\documentclass[12pt,preprint]{aastex}

\def\kms{\mbox{$\rm kms^{-1}$}}
\def\sec{$^{\prime\prime}$~}
\def\deg{$^\circ$}
\def\fc{$f_{mc}$}

\begin{document}

\title{CENTRAL MASS CONCENTRATION AND BAR DISSOLUTION IN NEARBY SPIRAL GALAXIES} 

\author{Mousumi Das\altaffilmark{1}, Peter J. Teuben\altaffilmark{1}, 
Stuart N. Vogel\altaffilmark{1}, Michael W. Regan\altaffilmark{2}, 
Kartik Sheth\altaffilmark{1,3}, Andrew I. Harris\altaffilmark{1}, 
William H. Jefferys\altaffilmark{4}}

\affil{\altaffilmark{1}Department of Astronomy, University of Maryland, 
College Park, MD 20742 }
\affil{\altaffilmark{2}Space Telescope Institute, 3700 San Martin Drive, 
Baltimore, MD 21218}
\affil{\altaffilmark{3}Department of Astronomy, California Institute of 
Technology, Pasadena, CA}
\affil{\altaffilmark{4}Department of Astronomy, University of Texas at Austin,
Austin, TX  78712}

\email{mousumi@astro.umd.edu}

\begin{abstract}

We use data from the BIMA Survey of Nearby Galaxies (SONG) to investigate
the relationship between ellipticity and central mass concentration in
barred spirals. Existing simulations predict that bar ellipticity
decreases as inflowing mass driven by the bar accumulates in the
central regions, ultimately
destroying the bar. Using the ratio of the bulge mass to the mass within
the bar radius as an estimate of the central mass concentration, we obtain
dynamical mass estimates from SONG CO 1-0 rotation curve data.  We find an
inverse correlation between bar ellipticity and central mass
concentration, consistent with simulations of bar dissolution.

\end{abstract}

Subject headings: galaxies:spiral --- galaxies:ISM --- galaxies:evolution --- \\
galaxies:structure --- interstellar:molecules ---  interstellar:kinematics and dynamics

\section{INTRODUCTION}

Bars exert gravitational torques on the gas in the disks of spiral
galaxies, resulting in gas inflow towards the center (Quillen et al.
1995; Regan, Vogel, \& Teuben 1997; Regan, Sheth, \& Vogel 1999).
This results in a significant increase in the gas mass in the center
of a bar (Sakamoto et al. 1999; Sheth 2001), often leading to
increased star formation and even starburst activity in the nucleus
(Ho, Filippenko, \& Sargent, 1997; Jogee, Kenney, \& Smith 1999).
Simulations predict that the increased mass concentration may affect
the bar structure and even dissolve the bar itself (Kormendy 1982;
Hasan, \& Norman 1990; Friedli, \& Pfenniger 1991; Friedli \& Benz
1993; Hasan, Pfenniger, \& Norman 1993; Norman, Sellwood, \& Hasan
1996).  In this paper we investigate whether there is observational
evidence for the change in bar shape with mass concentration in the
centers of spiral galaxies.

We have used the photometric data of a sample of 13 barred galaxies
from the BIMA Survey of Nearby Galaxies (SONG) to determine the bar
structure and the CO(1-0) rotation curves to derive central mass
concentrations.  We use the bar ellipticity $(1-b/a)$, where $a$ is
the semi-major axis and $b$ the semi-minor axis, to be a measure of
the bar structure. We define the central mass concentration `\fc'~ as
the ratio of the dynamical mass within the bulge to that within the
bar radius. The bulge is the most physically distinct region in the
galaxy center and easier to measure than other length scales such as
core radius which is used to define central mass concentration
in numerical studies (e.g.
Norman et al.  1996). We discuss the justification for using the bulge
mass in more detail in \S 5.  To determine \fc, we have used the
rotation curves derived from the CO J=1-0 emission in the galaxies.
CO rotation curves were used because CO traces the kinematics of cold
molecular gas, which moves along closed orbits in the plane of a
galaxy and hence is a good tracer of the dynamical mass distribution
in galaxies.  In \S 2 we describe the galaxy sample and the
observational data used in the analysis. In \S 3 we discuss how we
derived bar ellipticities and in \S 4 we determine \fc\ in our sample
of galaxies. The statistical analysis is presented in \S 5 and we
discuss the significance of the results in \S 6. We list our
conclusions in \S 7.

\section{GALAXY SAMPLE}

Our galaxy sample was taken from BIMA SONG; in this survey the centers
and inner disks of 44 nearby galaxies were imaged in the CO (1-0) line
(Regan et al. 2001; Helfer et al. 2002).  The galaxies were observed
using the Berkeley-Illinois-Maryland Association (BIMA) array at Hat
Creek (Welch et al. 1996).  Of the 44 galaxies, 29 are barred and 15
are unbarred.  The SONG database includes spatial-velocity cubes which
have resolutions of $\sim$4\sec- 6\sec and 4.1~\kms. More than half of
these galaxies had maps which included single-dish CO data taken with
the NRAO telescope (Helfer et al. 2002).  SONG also included a
parallel dataset of the near-infrared and optical images of the
galaxies. This was important for our study as we needed to determine
both the bar structure as well as the dynamical mass concentration in
our sample.

To determine the central mass concentration, we require in principle
only two velocity measurements in a galaxy, i.e. the rotation
velocities at the bulge and bar radii. But to be sure that the
velocities measured at these radii are not anomalous, we required good
velocity coverage for the CO emission over a significant portion of
the inner disk of these galaxies. Thus our sample size was limited by
the gas distribution to a subsample of 13 barred galaxies from the
BIMA SONG database (Table 1).

\section{BAR SIZE AND ELLIPTICITY}

Bar morphology has an important effect on the gas inflow and star
formation in galaxies (Martinet \& Friedli 1997, Aguerri 1999). Bar
strength can be quantified either by measuring the bar axis ratio
$b/a$ (Martin 1995; Regan \& Elmegreen 1997; Chapelon, Contini, \&
Davoust 1999), or alternatively by determining the maximum of the
ratio of the tangential force to the mean non-axisymmetric radial
force $(Q_b)$ in the bar (Buta \& Block 2001). Recent studies have
shown that bar ellipticity is roughly proportional to $Q_b$ (Laurikainen, Salo
\& Rautiainen 2002), and so ellipticity appears to be a good measure
of the bar strength. In our study we use optical and near-IR images
(R, I, and K bands) to determine the bar ellipticity, which we assume
provides a reasonable estimate of the bar strength.  The K and I bands
generally trace the old stellar population in galaxies and thus follow
the galactic potential.  We used K band images for NGC 3627 (Regan \&
Elmegreen 1997) and NGC 6946 (Regan \& Vogel 1995), and I band images for NGC
2903, NGC 3351, NGC 4303, NGC 4569, NGC 5248, and NGC 5457.  R band is
not as good a tracer of old stars but was all that was available for
NGC 3726.  For NGC 3184, NGC 3521, NGC 4321 and NGC 5005 we used
near-infrared (K band) images from the 2MASS survey (Jarrett et al. 2000).

The bar was identified from the isophotes in the optical/IR image. The
isophotes were traced using the $\tt{ellipse}$ task in IRAF, which is
based on the photometric technique developed by Jedrzejewski (1987).
The bar-defining isophote was assumed to be where the position angle
of the elliptical isophotes change direction and start tracing the
disk of the galaxy (Elmegreen et al. 1996; Laurikainen \& Salo 2000).
This isophote was used to determine the position angle and semimajor
axis length of the bar. We also determined the intensity profile of
the optical or near-IR emission along the bar axis and perpendicular
to it, using the IRAF task $\tt{pvector}$.  All the profiles have a
characteristic peak at the center due to the bulge and usually a flat
portion which represents the bar. Such profiles have been used in
previous studies to determine bar sizes and ellipticities (Elmegreen
et al. 1996; Regan \& Elmegreen 1997). We have assumed that the bulge
radius is the distance along the bar where the central peak appears to
end and the bar profile becomes prominent.  In some cases, for example
in NGC 4321, there appears to be a double bar since there are two
distinct flat portions in the profile (e.g. Knapen et al. 1995).
Double bars are evident both in the isophotes and in the intensity
profiles. In such cases, we had to carefully examine the profile in
order to distinguish the bulge from the rest of the bar.

Bright stars and dust can interfere with the isophote fitting
procedure; so to minimize this effect, we fitted the bar-defining
isophotes using the software suite NEMO (Teuben 1995).  To determine
deprojected values of the bar parameters, the bar was treated as a
two-dimensional ellipse and then projected onto the plane of the
galaxy. The parameters involved in the deprojection are the
inclination angle of the galaxy and the angle between the bar major
axis and galaxy axis. The parameters assumed for the galaxies and the
observed bar ellipticities are listed in Table 1 and the deprojected
values in Table 2.

\section{CENTRAL MASS CONCENTRATION FROM ROTATION CURVES}

We have used the rotation curves derived from the CO(1-0) emission
line observations of BIMA SONG to determine \fc\ for our sample of
barred galaxies. As noted earlier, the CO line traces the molecular
gas distribution in the bar and should be a good tracer of the bar
potential because it is cold, and hence should settle along closed
orbits in the plane of the galaxy. We used the spatial-velocity data
cube derived from the spectral line observations to determine the
zeroth and first order moments of the intensity distribution, i.e. the
integrated CO intensity maps and the mean-line-of sight velocity maps
for these galaxies.

A detailed description of the methods involved in deriving these
rotation curves has been discussed elsewhere 
(Begeman 1987; Teuben et al. 2002, in preparation). We
summarize the procedure here.  We used a two step process to obtain
the CO velocity fields. First, we generated a mask cube by smoothing
the cube at each velocity channel with a $20$\arcsec ~gaussian; only
those pixels with emission brighter than the 3$\sigma$ level in this
smoothed cube were allowed by the mask. Then, we further masked the
data by including only pixels with brightness above the 2$\sigma$
level in the unsmoothed cube. Moment maps at the full resolution of
the data were then generated in the standard way using these two
masks.  The rotation curves were determined from the velocity fields
using the NEMO task $\tt{rotcur}$, which is based on the tilted-ring
model fitting method developed for HI rotation curves (Begeman 1989).
The parameters for the rings are the inclination angle, the position
angle, the systemic velocity of the galaxy and the rotation center of
the galaxy.  The dynamical center for a galaxy was assumed to be the
brightness centroid derived from the K, I or R band images; this
involved examining the intensity contours for the central region and
taking the dynamical center to be where the contours converged to a
maximum intensity.  Both the galaxy center and inclination angle were
kept constant for all the rings. An initial estimate of the position
angle and the systemic velocity of a galaxy was taken from NED,
but it was clear from examination of the rotation curve
fits and residuals that the literature values needed to be changed for
many galaxies in our sample.  We iteratively changed the position
angle and the systemic velocity of a galaxy until the errors for the
velocity fits in the individual rings in the rotation curve were a
minimum. The new values were all close to the literature values. The
new improved parameters that we adopted are listed in Table 1. As
usual for the tilted-ring method, a rotation velocity is derived for
each ring, weighting pixels with respect to the major axis of the
galaxy.  The width of a ring was set to the mean synthesized
beamwidth; for Nyquist sampling, we used rings spaced at half beam
intervals.

The rotational velocities at the bulge radii ($R_{blg}$) and at the
bar ends ($R_{bar}$) were interpolated from the rotation curves. We
derive the mass concentration as
\fc=${v_{blg}}^{2}R_{blg}/{v_{bar}}^{2}R_{bar}$, where $v_{blg}$ and
$v_{bar}$ are the rotational velocities at the bulge radius and at the
bar semi-major axis length respectively. This relation does not
include the effects of bulge and disk geometry which will be different
for each galaxy. It also assumes that the magnitude of elliptical
streaming is not significant at the bulge radius or bar ends. This is
a reasonable assumption in the bulge where the potential is fairly
axisymmetric, and also at the bar ends where the disk potential begins
to be more important than the bar. We discuss evidence supporting this
assumption in the next section.  The values of \fc\ derived for our
sample of bars are shown in Table 2. Also shown are the dynamical
masses within the bulge and bar of the galaxies.

\section{CORRELATION OF BAR ELLIPTICITY AND CENTRAL MASS CONCENTRATION} 

Figure 1 shows the deprojected ellipticities in the plane of the
galaxies, plotted against the central mass concentration \fc.  The
errors for both axes have been calculated using the standard error
propagation equation, based on the uncertainties in the observed
quantities (Bevington \& Robinson 1992).  The error along the
ellipticity axis includes a coefficient due to the deprojection of the
bar onto the plane of the galaxy.  It is clear even from just visual
inspection that there is a correlation between bar ellipticities and
\fc\ in the galaxies. However, for a more quantitative estimate of the
correlation, we have determined a linear correlation coefficient for
the sample using two different methods.

An accurate estimate of the correlation coefficient should include the
errors on both axes, which leads to a weighted correlation
coefficient; but this is difficult to obtain in practice (Feigelson \&
Babu 1992).  So we used a simple Monte Carlo simulation to determine a
mean weighted correlation coefficient.  We generated 20,000 linear
fits to the points in the sample where each line randomly sampled the
errors on both axes.  The mean correlation coefficient $\bar{r}$ of
all the fits was assumed to be the weighted correlation coefficient
for the sample. We obtained a value of $\bar{r}=-0.86$, which
indicates a significant correlation. It is also important to determine
the p-value $P_r$ corresponding to such a linear correlation
coefficient (Bevington \& Robinson 1992). For $\bar{r}=-0.86$, $P_r~<~0.001$ so
the probability that they are from a random sample is small. The
second approach was to use the Bayesian model fitting method (Sivia
1996; Loredo 1990). This technique determines the probability of a
correlation assuming the errors on both axes to be independent and
Gaussian. The resulting integral was solved using the
Markov Chain Monte Carlo technique (Casella \& George 1992).  This
more rigorous method, which used conventional priors for the analysis,
resulted in a mean correlation coefficient of
$\bar{r}=-0.75\pm0.1$ and the posterior probability of the
uncorrelated model is 0.02. Thus both methods confirm the significant
correlation evident by eye.

From Figure 1 it is clear that the three galaxies NGC 5005, NGC 3184
and NGC 3521, which have ellipticity less than 0.35, are important for
the correlation. So we have examined these galaxies in closer detail
and compared their bar properties and rotation curves with other
observations in the literature.

{\it NGC 5005} : In K band, this galaxy appears to have a strong bulge and
a fairly round bar.  But there are two features that at first glance
appear inconsistent with this. First, there are two straight, structures
parallel to the length of the bar; they make the bar appear more
elliptical than our estimate from isophote fitting. On closer inspection
they appear offset from one another, rather like dust lanes. However, dust
emission should not be prominent in the K band so these structures cannot
represent dust lanes; their origin is thus not clear and requires further
investigation. An alternative estimate of how round a bar is can be
obtained from the bar strength parameter $Q_b$. E.Laurikainen (private
communication) found a low value of $Q_b=0.16$ for the bar strength in
this galaxy.  The second feature is the velocity field, which exhibits
prominent shocks that are typically considered a feature of strong bars,
not weaker bars such as NGC 5005.  However, using the Piner, Stone, \&\
Teuben 1995 hydrodynamical code, we simulated a galaxy with a bar axis
ratio similar to NGC 5005 and found strong shocks similar to that
observed.  Thus we conclude that despite the strong shocks,
a weak bar with $Q_b=0.16\pm0.5$ is possible.
 
The rotation curve of this galaxy is affected by beam smearing, which
is made worse by the high inclination angle of this galaxy ($61.4^{\circ}$).
Also, since the bar is fairly closely aligned with the major axis in
this galaxy, the PV plot is affected by the elliptical streaming of
gas in the bar. This explains why the rotation curve has a velocity at
the bulge radius lower than that seen in the PV diagram; this was also
evident when we compared our rotation curve with the PV plot of
Sakamoto, Baker and Scoville (2000). However, the beam smearing effect
is important for mainly the inner 2-3 beamwidths, which is well within
the bulge radius for this galaxy. Also, the effect of elliptical
streaming is considerably reduced in deriving the rotation curve
because velocities are azimuthally averaged over annuli. We thus
believe our isophote measurement of the bar ellipticity and the
rotation curve determination of \fc\ are both reasonable estimates for
NGC 5005.

{\it NGC 3184} : This galaxy has both a large bulge and a round bar.
The bar is not prominent in the K band image. However the CO distribution 
shows the classic response of gas in a barred potential, with trailing 
spiral arms emerging from the ends of the bar.
The bulge is a little over half the bar size and is very
bright in the near-IR, which may indicate a large mass concentration
in the center of the galaxy. 

{\it NGC 3521} : The ellipticity of this galaxy is so low, it might be 
questionable whether it is in fact a barred galaxy. But first we
note that this galaxy is classified as a barred galaxy in the
RC3 catalogue (Table 1). However, the bar is not easy to distinguish. This
is partly because of the low ellipticity of the bar but also because
of the high inclination of the galaxy (62\deg). We found evidence for
the bar in the K band photometry and also in the intensity profile
along the major axis of the galaxy. Also, Zeilinger et al.  (2001)
find evidence for the bar both in their R band photometry and in the
stellar kinematics derived from the long slit spectra obtained along
different axes in the galaxy. This leads us to believe that there is a
rather round bar in this galaxy. The bulge is very bright and very large;
it it over 3/4 the size of the bar and this indicates that it may also
be very massive. 

It therefore appears that the three galaxies with low measured
ellipticities are indeed barred galaxies with relatively low
ellipticity and with high central mass concentrations. Nonetheless, it
will be important to confirm the trend identified here using a larger
sample of galaxies.

The last issue which should be discussed is the effect of elliptical
streaming in the bar, due to gas moving on $x_{1}$ and $x_{2}$ orbits. 
The $x_{1}$ orbits are aligned along
the bar within corotation radius and the $x_{2}$ orbits are elongated 
perpendicular to the length of the bar (Contopoulos \& Papayannopoulos 1980; 
Binney \& Tremaine 1987).  We have tried to minimize the effect of elliptical
streaming by measuring velocities at the bulge and bar radii. We 
assume that we are measuring the $x_2$ orbits at the edge of the bulge
and the $x_1$ orbits at the edge of the bar.  For a bar aligned with
the major axis of the galaxy, $v_{blg}$ is measured at the pericenter
of its $x_2$ orbit and therefore larger than a circular orbit
consistent with the mass interior to that radius. Equally so,
$v_{bar}$ is measured at the apocenter of its $x_1$ orbit and
therefore smaller then the corresponding circular orbit.  Therefore
\fc\ would be overestimated from a value consistent with the mass
distribution.  Conversely for bars aligned along the minor axis \fc\ 
would be underestimated. We did not see such a trend in Figure 1. We
should also note that the rotation curve is derived from a two
dimensional velocity field which will tend to average out this effect.

\section{DISCUSSION}

Figure 1 shows that galaxies which have more mass concentrated in
their bulges have rounder bars.  However, the existence of a
correlation does not necessarily imply a causal relationship.  For
example, it might be that some other parameter used to calculate the
masses is more revelant.  In particular, the central mass
concentration is calculated as
\fc =${v_{blg}}^{2}R_{blg}/{v_{bar}}^{2}R_{bar}$; it might be that the
relative bulge size $l_{c}=R_{blg}/R_{bar}$ is more relevant than the
dynamical mass concentration.To investigate this, we plotted
ellipticities against $l_{c}$ for 25 barred SONG galaxies (Figure 2).
Comparison with Figure 1 shows that the correlation of ellipticity
with $l_{c}$ is not as good as with $f_{mc}$; the linear correlation
coefficient (unweighted) is $-$0.56, significantly less than for \fc.
We therefore conclude that \fc\ is more likely than $l_{c}$ to be
relevant for the central mass concentration.  The points in Figure 2
are coded to represent the 3 Hubble types i.e.  early, intermediate
and late type barred galaxies. Early type galaxies have large, bright
bulges whereas late type galaxies appear to have less prominent bulges
(Hubble 1926). In Figure 2, the different types are dispersed over the
whole plot; there does not appear to be any correlation of bar
ellipticity with Hubble type.  This may because the Hubble sequence is
based on the optical brightness of bulges which may not always be a
good measure of the bulge mass. We thus conclude that bar ellipticity
is less correlated with bulge size or Hubble type than with central
mass concentration. Thus Figure 1 indicates that bar ellipticity,
which approximately measures bar strength, decreases with mass
concentration in the bar.

The correlation is perhaps not suprising, since a spherical mass
concentration in the center of a bar will tend to decrease the
non-axisymmetric effect of the bar.  Nonetheless, it is worth pointing
out that this correlation is predicted by secular evolution models, as
we now discuss.
Simulations show that bars drive gas inwards, resulting in central
star formation and a consequent buildup of the central mass. The
central mass concentration affects the stability of the bar supporting
$x_1$ orbits so that the bar finally dissolves leading to the
transformation of a barred spiral to an unbarred one (Friedli \& Benz
1993). Other simulations indicate that it may be the massive core that
affects the $x_1$ orbits, causing them to become stochastic and
leading to the dissolution of the bar (Hasan, Pfenniger \& Norman
1993, Norman et al. 1996; Hozumi \& Hernquist 2001).  These models
also explain the lack of bars with low ellipticity because once
\fc\ reaches a certain range, the bar evolves very rapidly. It has also
been suggested that bar dissolution may be due to the scattering of
stars by a large mass concentration in the center of a galaxy (Norman,
May, \& van Albada 1985; Gerhard \& Binney 1985) or even a prolate
halo (Ideta and Hozumi 2000).

We have, however, measured the bulge concentration \fc\  and not the
core mass.  Nonetheless, \fc\ may be a useful measure of the core mass
concentration.  Observations of the black hole or core masses in
galaxies indicate they are well correlated with the bulge mass
(Magorrian et al. 1998; Richstone et al. 1998; Ferrarese \& Merritt
2000; Gebhardt et al. 2001).  This means that though \fc\  does not
directly measure a nuclear mass concentration, it may be a reasonable
measure of the effect of a massive core on the overall bar morphology.
Table 2 shows the dynamical masses in the bulge and bar calculated
using the approximation, $M_{dyn}\sim{rv^{2}}/G$.  Assuming galaxies
have a dynamical mass lying in the range $10^{11} - 10^{12}
M_{\odot}$, some of the galaxies in our sample have bulge masses that
are a few percent of the total galaxy mass.  This may be large enough
to affect the $x_1$ orbits in these galaxies and further increase
could lead to bar dissolution, which will leave behind a spheroidal
bulge in the center of the galaxy. This is but one of several
processes such as accretion or bending instabilities that lead to
bulge formation (e.g. Raha et al. 1991; Carlberg 1992).  This should
not, however, prevent galaxies from reforming bars again if the disk
is cool enough; this must evidently be the case since a significant
fraction of all spiral galaxies are unbarred.  Thus bar formation,
dissolution and bulge formation may be an ongoing evolutionary process
in galaxies (Bournaud \& Combes 2002).

We end the discussion on a note of caution in using Figure 1 in
support of bar dissolution models. First, a larger sample is required
to confirm the apparent correlation. Second, even though ellipticity
and central mass concentration are correlated, this of course does not
require a causal connection between the two.  Third, if bars dissolve
significantly faster than a Hubble time, we need to understand how
they reform since there is a significant fraction of barred galaxies.

\section{CONCLUSIONS}

We have used the BIMA SONG survey data to determine ellipticities and
mass concentrations in the centers of nearby barred galaxies. We have
used optical or near-infrared images to determine bar shapes and the
CO (1-0) rotation curves to derive dynamical masses in the bulge and
bar regions of the galaxies.  \\ 1) We find an apparent correlation
between the bar ellipticity and the central mass concentration. For
our sample of 13 galaxies a conservative analysis yields a correlation
coefficient of $\sim -0.8$. The probability that the parent sample is
uncorrelated is 0.012, which indicates that it is a statistically
significant correlation.  \\ 2) The correlation suggests that bar
structure is affected by the dynamical mass concentration in the
bulge. This may provide evidence that bars evolve as gas flows inwards
and mass accumulates in their centers, indicating that the mass
concentration affects the bar structure and may eventually dissolve
the bar.

\acknowledgments

M.D. thanks S.~Jogee, E.~Laurikainen, W.~W.~Zeilinger and K.~Sakamoto
for useful discussions; we thank the SONG team for providing the
entire dataset used in this paper.  This work is partially supported
by NSF AST-9981289.  This research has made use of the NASA/ IPAC
Infrared Science Archive, which is operated by the Jet Propulsion
Laboratory, California Institute of Technology, under contract with
the National Aeronautics and Space Administration.

\newpage

\centerline{\bf FIGURE CAPTIONS}

\noindent Figure 1 Deprojected bar ellipticity $(1-b/a)_d$ plotted
against the central mass concentration (\fc) in the bar.  Error bars
are $1\sigma$ for each axis. The 3 galaxies with the largest central
mass concentration, \fc, are in decreasing order NGC 3521, NGC 3184
and NGC 5005 respectively.

\noindent Figure 2 Deprojected bar ellipticity plotted against the ratio of 
bulge to bar radii $(l_c)$ for 25 SONG galaxies. The open squares
represent early type galaxies, the open triangles represent
intermediate galaxies and the stars represent late type galaxies.

\clearpage
{\scriptsize
\begin{deluxetable}{lcccccccccc}
\tablenum{1}
\tablewidth{0pt}
\tablecaption{Observed Bar Parameters}
\tablehead{
\colhead{Galaxy} & \colhead{Type} & \colhead{$V_{sys}$} 
& \colhead{Distance} & \colhead{Inclination}
& \colhead{PA} & \colhead{PA} & \colhead{Bar}
& \colhead{Bulge} & \colhead{Ellipticity} & \\
& \colhead{(RC3)$^{1}$} & \colhead{($\rm km s^{-1}$)} & \colhead{(Mpc)} & \colhead{}
& \colhead{(galaxy)} & \colhead{(bar)} & \colhead{radius(\sec)} 
& \colhead{radius(\sec)} & \colhead{($1-b/a$)}
}
\startdata
NGC 2903 & SAB(rs)bc & 556 &  7.3   & 61.4\deg  & 17\deg  & 24\deg  & 67 & 17 & 0.79\\ 
NGC 3184 & SAB(rs)cd & 592 & 8.7   & 21.1\deg & 180\deg & 62\deg  & 21 & 13  & 0.21\\
NGC 3351 & SB(r)b    & 774 & 10.1 & 47.5\deg & 13\deg  &  113\deg & 54 & 19 & 0.42\\ 
NGC 3521 & SAB(rs)bc & 805 & 7.2   & 62.1\deg & 163\deg & 162.6\deg & 25 & 20 & 0.52\\
NGC 3627 & SAB(s)b   & 727 & 11.1 & 62.8\deg  & 173\deg & 161\deg & 49 & 26 & 0.68\\ 
NGC 3726 & SAB(r)c   & 861 & 11.7  & 46.2\deg & 10\deg  & 32\deg  & 32 & 12 & 0.69\\
NGC 4303 & SAB(rs)bc & 1562 & 15.2  & 27\deg & 138\deg & 2\deg   & 47  & 16 & 0.53\\
NGC 4321 & SAB(s)bc  & 1566 & 16.1  & 31.7\deg & 153\deg & 108\deg & 54 & 20 & 0.51\\
NGC 4569 & SAB(rs)ab & -221 & 16.8  & 62.8\deg  & 23\deg  & 16\deg  & 65 & 24 & 0.71\\ 
NGC 5005 & SAB(rs)bc & 950  & 21.3  & 61.4\deg  & 65\deg  & 74\deg  & 37 & 19 & 0.63\\
NGC 5248 & SAB(rs)bc & 1158 & 22.7  & 43.6\deg & 106\deg & 127\deg & 80 & 16 & 0.50\\ 
NGC 5457 & SAB(rs)cd & 258  & 6.5  & 21.1\deg & 35\deg  & 80\deg  & 49 & 13 & 0.38\\ 
NGC 6946 & SAB(rs)cd & 48   & 6.4   & 31.7\deg & 64\deg  & 23\deg  & 63 & 15 & 0.44\\ 
\enddata
\tablerefs{
(1) de Vaucouleurs et al. 1995}
\end{deluxetable}
}

\clearpage

\begin{deluxetable}{lccccc}
\tablenum{2}
\tablewidth{0pt}
\tablecaption{Derived Bar Parameters}
\tablehead{
\colhead{Galaxy} & \colhead{Dynamical Mass in} & \colhead{Dynamical Mass in}
& \colhead{Central Mass } & \colhead{Deprojected}\nl
\colhead{ } & \colhead{Bulge ($\rm 10^{8}M_{\odot}$)} & \colhead{bar ($\rm 10^{8}M_{\odot}$)}
& \colhead{Concentration} & \colhead{Ellipticity}
}
\startdata
NGC 2903 & 8   & 170  & 0.05 & 0.59 \\ 
NGC 3184 & 4   &  10  & 0.45 & 0.24 \\ 
NGC 3351 & 18  & 306  & 0.06 & 0.60 \\ 
NGC 3521 & 46  & 69   & 0.67 & 0.04 \\ 
NGC 3627 & 44  & 245  & 0.18 & 0.43 \\ 
NGC 3726 & 4   & 61   & 0.07 & 0.62 \\ 
NGC 4303 & 27  & 218  & 0.12 & 0.53 \\ 
NGC 4321 & 58  & 374  & 0.16 & 0.52 \\ 
NGC 4569 & 62  & 448  & 0.14 & 0.42 \\ 
NGC 5005 & 246 & 760  & 0.32 & 0.33 \\
NGC 5248 & 108 & 596  & 0.18 & 0.40 \\  
NGC 5457 & 12  & 77   & 0.15 & 0.38 \\  
NGC 6946 & 23  & 173  & 0.13 & 0.45 \\  

\enddata
\end{deluxetable}

\clearpage

\plotone{f1.eps}
\plotone{f2.eps}


\newpage


\begin{references}
\reference {} Aguerri, J.A.L. 1999, A\&A, 351, 43
\reference {} Begeman, K.G. 1987, PhD. Thesis, University of Gronigen
\reference {} Begeman, K.G. 1989, A\&A, 223, 47
\reference {} Bevington, P.R., \& Robinson, D.K. 1992, 2nd edition, McGraw-Hill, Inc
\reference {} Binney, J., \& Tremaine, S. 1987, Princeton University Press 
\reference {} Bournaud, F., \& Combes, F. 2002, A\&A, astroph/0206273
\reference {} Buta, R., \& Block, D.L. 2001, ApJ, 550, 243
\reference {} Carlberg, R.G. 1999, in The Formation of Galactic Bulges, ed. C.M. Carollo, H.C. Ferguson, R.F.G.Wyse,
\reference {} Casella, G., \& George, E.I. 1992, American Statistician, 46, 167 
(Cambridge University Press, Cambridge, U.K.), 64
\reference {} Chapelon, S., Contini, T., Davoust, E. 1999, A\&A, 345, 81 
\reference {} Contopoulos, G., \& Papayannopoulos, T. 1980, A\&A, 92, 33
\reference {} de Vaucouleurs, G., de Vaucouleurs, A., Corwin, H.G., Buta, R.J., Paturel, G., Fouque, P. 1995, Third Reference Catalogue of Bright Galaxies (RC3), Springer-Verlag, New York  
\reference {} Elmegreen, B.G., Elmegreen D.M., Chromey, F.R., Hasselbacher, D.A., Bissel, B.A., 1996, AJ, 111, 2233
\reference {} Feigelson, E.D., \& Babu, G.J. 1992, ApJ, 397, 55
\reference {} Ferrarese, L., \& Merritt, D. 2000, ApJL, 539, L9
\reference {} Friedli, D., \& Pfenniger, D 1991, in IAU Symp. 146, Dynamics of Galaxies and Their Molecular Cloud Distributions, ed F.Combes, \& F.Casoli (Dordrecht: Reidel) 362 
\reference {} Friedli, D., \& Benz, W. 1993, A\&A, 268, 65
\reference {} Gebhardt, K. et al. 2000, ApJL, 539, L13 
\reference {} Gerhard, O.E., \& Binney, J. 1993, A\&A, 268, 65   
\reference {} Hasan, H., \& Norman, C. 1990, ApJ, 361, 69
\reference {} Hasan, H., Pfenniger, D., \& Norman, C. 1993, ApJ, 409, 91
\reference {} Helfer, T.T. 2002, submitted  
\reference {} Ho, L.C., Filippenko, A.V., \& Sargent, W.L.W. 1999, ApJ, 487, 591
\reference {} Hozumi, S., \& Hernquist, L. 2002, ApJL, submitted 
\reference {} Hubble, E. 1926, ApJ, 64, 321
\reference {} Ideta, M., \& Hozumi, S. 2000, ApJ, L91
\reference {} Jarrett, T.H., Chester, T., Cutri, R., Schneider, S., Skrutskie, M., \& Huchra, J.P. 2000, AJ, 119, 2498
\reference {} Jedrzejewski, R. I. 1987, MNRAS, 226, 747
\reference {} Jogee, S., Kenney, J.D.P., \& Smith, B.J. 1999, 526, 665
\reference {} Knapen, J.H., Beckman, J.E., Shlosman, I., Peletier, R.F., Heller, C.H., de Jong, R.S. 
1995, ApJ, 443, L73
\reference {} Kormendy, J. 1982, in Proc. Saas Fee Course, No. 12, Morphology and Dynamics of Galaxies, ed. L.Martinet \& M.Mayor (Sauverny : Geneva Obs.), 113
\reference {} Laurikainen, E., \& Salo, H. 2000, A\&AS, 141, 103
\reference {} Laurikainen, E., Salo, H. \& Rautiainen, P. 2002, MNRAS, 331, 880L 
\reference {} Loredo, T.J. 1990, Maximum Entropy and Bayesian Methods, eds. Fougere, P.F., 
Dordrecht, Kluwer, p81
\reference {} Magorrian et al. 1998, AJ, 115, 2285 
\reference {} Martin, P. 1995, AJ, 109, 2428
\reference {} Martinet, L., \& Friedli, D. 1997, A\&A, 323, 363 
\reference {} Norman, C., May, A., \& van Albada, T. 1985, ApJ, 296, 20
\reference {} Norman, C., Sellwood, J.A, \& Hasan, H. 1996, ApJ, 462, 114  
\reference {} Piner, B. G., Stone, J. M., \& Teuben, P. J. 1995, ApJ, 449, 508
\reference {} Quillen, A.C., Frogel, J.A., Kenney, J.D.P., Pogge, R.W., Depoy, D.L. 1995, ApJ, 441, 549 
\reference {} Raha, N., Sellwood, J.A., James, R.A., Kahn, F.D. 1991, Nature, 352, 411
\reference {} Regan, M. W., \& Vogel, S.N. 1995, ApJ, 452, L21
\reference {} Regan, M. W., Vogel, S.N., Teuben, P.J. 1997, ApJ, 482, L143
\reference {} Regan, M. W., \& Elmegreen, D. B. 1997, AJ, 114, 965
\reference {} Regan, M. W., Sheth, K., \& Vogel 1999, ApJ, 526, 97
\reference {} Regan, M. W. et al. 2001, ApJ, 561, 218  
\reference {} Richstone, D. et al. 1998, Nature, 395, A14 
\reference {} Sakamoto, K., Okumura, S.K., Ishizuki, S., \& Scoville, N.Z. 1999, ApJ, 525, 691
\reference {} Sakamoto, K., Baker, A.J., \& Scoville, N.Z. 2000, ApJ, 533, 149 
\reference {} Sellwood, J.A., \& Moore, E.M. 1999, ApJ, 510, 125
\reference {} Sheth, K. 2001, PhD. Thesis, University of Maryland
\reference {} Sivia, D.S. 1996 Data Analysis : A Bayesian Tutorial, New York, Oxford University Press
\reference {} Teuben, P.J. 1995, ADASS IV, ASP Conference Series, Vol 77, eds. Shaw, R.A., Payne, H.E., Hayes, J.J.E., pg. 398   
\reference {} Welch, W.J. et al. 1996, PASP, 108, 93
\reference {} Zeilinger, W.W., Vega Beltran, J.C., Rozas, M., beckman, J.E., Pizella, A., Corsini, E.M., Bertola, F. 2001, Ap\&SS, 276, 643

\end{references}
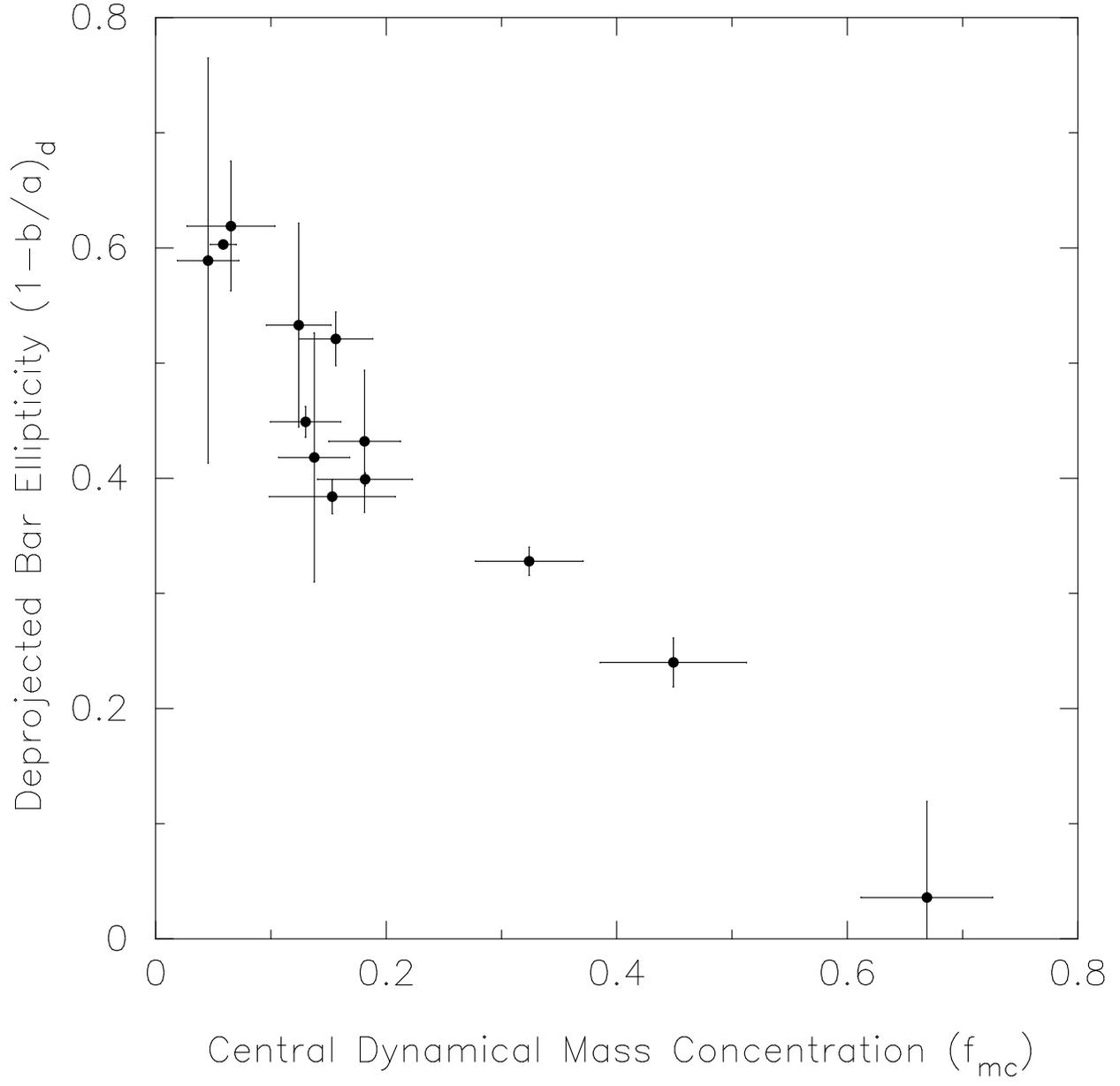
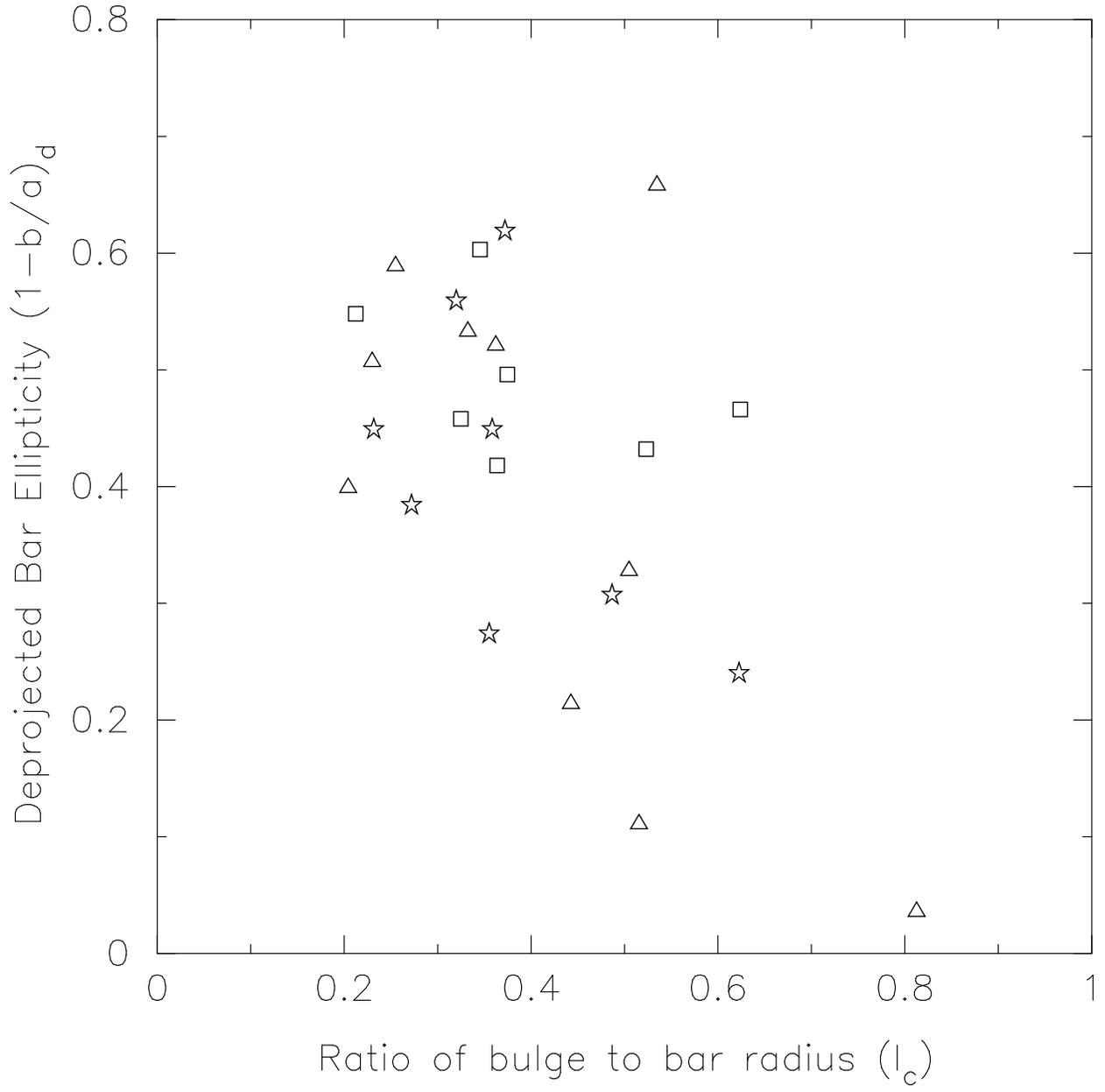
\end{document}